\documentclass[sigconf]{acmart}

\usepackage{tikz}
\usetikzlibrary{arrows.meta,positioning}
\usepackage{booktabs}

\copyrightyear{2026}
\acmYear{2026}
\setcopyright{cc}
\setcctype{by}
\acmConference[ICMI Companion '26]{Companion of the INTERNATIONAL CONFERENCE ON MULTIMODAL INTERACTION}{October 05--09, 2026}{Napoli, Italy}
\acmBooktitle{Companion of the INTERNATIONAL CONFERENCE ON MULTIMODAL INTERACTION (ICMI Companion '26), October 05--09, 2026, Napoli, Italy}
\acmDOI{10.1145/3776591.3833866}
\acmISBN{979-8-4007-2319-3/2026/10}

\begin{document}

\title{From Blind Edits to Verified Repair: Building Trustworthy User-Side LLM Agents for Web Accessibility}

\author{Lily Bundgaard Wanscher}
\email{liwan21@student.sdu.dk}
\affiliation{%
  \institution{University of Southern Denmark}
  \city{Odense}
  \country{Denmark}}

\author{Markus Heidemann Lorensen}
\email{mlore22@student.sdu.dk}
\affiliation{%
  \institution{University of Southern Denmark}
  \city{Odense}
  \country{Denmark}}

\author{Mohammed Ammad Shafiq}
\email{mosha22@student.sdu.dk}
\affiliation{%
  \institution{University of Southern Denmark}
  \city{Odense}
  \country{Denmark}}

\author{Mahyar Tourchi Moghaddam}
\email{mtmo@mmmi.sdu.dk}
\affiliation{%
  \institution{University of Southern Denmark}
  \city{Odense}
  \country{Denmark}}

\author{Mina Alipour}
\email{mial@mmmi.sdu.dk}
\affiliation{%
  \institution{University of Southern Denmark}
  \city{Odense}
  \country{Denmark}}

\renewcommand{\shortauthors}{Wanscher et al.}

\begin{abstract}
Assistive agents that adapt web pages on the user's side, at the moment of browsing, could reach the accessibility failures that site authors leave unfixed, and large language models make such agents newly plausible. We contribute three building blocks toward that goal. The first is a complete, privacy-preserving browser agent: a Chrome extension that extracts a page's style sheets, condenses them to fit a local model's context window, asks the model for additive CSS addressing 18 metrics from WCAG and the W3C cognitive accessibility guidance, and injects the result reversibly into the live page. The second is a dual-condition protocol that measures harm as carefully as benefit, applied to six small open-weight models (7B to 14B) on ten violation-rich and ten highly accessible live sites. The diagnosis is sobering but precise: unverified generation improved and regressed pages at similar rates (24 improvements against 20 regressions across the 100 trials of the five models that produced injectable CSS), fixing typography while breaking perception-dependent properties. The third answers the diagnosis: a verified repair instrument pairing a trilingual seeded-violation benchmark with an audit-inject-verify loop that accepts a change only if violations strictly decrease, so regression on the automated checks is impossible by construction. In a real browser the instrument detects 57 of 57 seeded violations with no false positives and rejects 126 of 126 adversarially harmful candidates. All code, prompts, benchmark materials, aggregate data, and validation logs are released.
\end{abstract}



\maketitle

\section{Introduction}
The accessibility of the web is not improving. The 2026 WebAIM Million analysis of the top one million home pages detected WCAG~2 conformance failures on 95.9\% of them, an average of 56.1 distinct errors per page, with low contrast text alone appearing on 83.9\% of pages~\cite{webaim2026million}. Two decades of mature guidelines have not closed the gap~\cite{wcag21,wcag22}, and the same report suggests AI-assisted code generation may now be widening it~\cite{webaim2026million}.

Waiting for millions of site authors to remediate their pages is one strategy. A complementary strategy, with a long history in accessibility research, is to adapt content on the \emph{user's} side: transcoding proxies, collaborative scripting, and community metadata authoring all repair pages at the moment of consumption, without requiring cooperation from the site~\cite{richards2004broader,bigham2007accessmonkey,takagi2008social}. Large language models revive this vision in general form: if a model can read a page's style sheets and rewrite them, one browser agent could in principle adapt \emph{any} page to \emph{any} user's presentation needs, from dyslexia-friendly type to visible focus indicators. Because such an agent necessarily observes everything the user browses, running the model locally is attractive for privacy, which pushes toward small open-weight models on consumer hardware, and the browser extension is the natural delivery vehicle: it sits inside the user's session, sees the fully rendered page, and layers changes onto any site without proxies or per-site engineering. If this recipe worked, accessibility work would shift from millions of uncoordinated site authors to one piece of assistive software under the user's own control.

This paper asks whether that vision survives contact with the current generation of small local models. We built a minimal but complete instantiation: a Chrome extension that extracts the active tab's styling, condenses it into a token budget by filtering and ranking rules, prompts a locally hosted model (via Ollama~\cite{ollama2026}) for additive CSS addressing 18 metrics drawn from WCAG~2~\cite{wcag21,wcag22} and the W3C cognitive accessibility guidance~\cite{coga2021}, and injects whatever valid CSS it can recover from the response. We evaluated six open-weight models, 7B to 14B parameters across four vendors, on twenty live websites: ten violation-rich commercial sites and ten widely recognized accessible sites serving as a do-no-harm control, scoring all 18 metrics before and after each intervention.

We ask four questions. \textbf{RQ1 (Design):} can an LLM-mediated browser intervention work without site cooperation, on consumer hardware? \textbf{RQ2 (Scope):} which metrics can a CSS-only intervention plausibly address? \textbf{RQ3 (Impact):} how much does it improve, and how much does it harm, on violation-rich and on already accessible sites? \textbf{RQ4 (Models):} do small local models differ measurably? Two directional hypotheses were set at $\alpha = .05$: significant improvement on violation-rich sites, and only an insignificant share of new violations on accessible ones.

RQ1 turned out that the pipeline is straightforward, and its engineering lessons transfer directly to future systems. The answers to RQ3 and RQ4 are instructive rather than encouraging, and we report them exactly. Across the 100 planned trials of the five models that produced injectable CSS, 24 improved a site and 20 made one measurably worse; the sixth model produced no extractable CSS in any of its twenty trials; no test approaches significance. New violations were, if anything, \emph{more} frequent on the accessible control sites. We treat this diagnosis as a specification. The failures localize to one missing component, verification of each edit against the rendered page, so we built and validated that component: a seeded-violation benchmark and a verified repair loop whose accept rule makes regression on the automated checks impossible by construction. In a browser it recovers every seeded violation with no false positives, rejects every candidate from an adversarially harmful generator, and turns model evaluation from weeks of manual scoring into hours of unattended runs.

This paper contributes:
\begin{itemize}
  \item an open-source, privacy-preserving browser agent for LLM-driven CSS repair, with a context-budgeting scheme that fits megabytes of production CSS into a 10{,}000-token window (Section~\ref{sec:system});
  \item a dual-condition protocol pairing violation-rich sites with an accessible control set under an 18-metric rubric, applied across six models and 120 planned trials to show where unverified small-model repair succeeds and where it harms (Sections~\ref{sec:study} and \ref{sec:results});
  \item a validated verified-repair instrument: a trilingual seeded benchmark plus an audit-inject-verify loop with a do-no-harm guarantee on its checks (57/57 seeded violations detected, zero false positives, 100\% of harmful candidates rejected; Section~\ref{sec:instrument});
  \item an analysis of the failure modes and the design requirements they imply: perceptual grounding, verification, reversibility (Section~\ref{sec:discussion}).
\end{itemize}
A replication package with the extension source, prompts, rubric, benchmark, harness, summary data, and analysis scripts accompanies this paper \footnote{\url{https://figshare.com/s/baec720691de58f99f92}}.

\section{Related Work}

\subsection{Accessibility Guidelines and Their Evaluation}
The Web Content Accessibility Guidelines define the de facto conformance target for the web~\cite{wcag21,wcag22}, and the W3C's Cognitive and Learning Disabilities Accessibility Task Force supplements them with guidance, such as content organization and clear typography, that goes beyond testable success criteria~\cite{coga2021}. Evaluating conformance is itself a research area. Automated checkers catch only a fraction of real barriers, and sole reliance on them measurably harms evaluation quality~\cite{vigo2013benchmarking}. Conformance and experienced accessibility also diverge: Power et~al.\ found that only about half of the problems blind users encountered were covered by WCAG~2.0 at all~\cite{power2012guidelines}, and Petrie and Kheir showed that accessibility and usability problems overlap only partially~\cite{petrie2007relationship}. Recent work argues for hybrid frameworks combining automated checks, semantic analysis, and human assessment~\cite{ara2025inclusive}, while comparative audits keep documenting poor conformance in practice~\cite{fakrudeen2025evaluation}. Strikingly, the same six error categories have dominated the WebAIM Million for seven consecutive years and account for 96\% of detected failures~\cite{webaim2026million}; half are presentation-level properties that CSS controls, which is what makes a style-level repair agent worth attempting. We take the lesson to heart: our rubric is grounded in WCAG and COGA, scored by a human, and treated as a proxy that bounds, not equals, user experience.

\subsection{User-Side Adaptation}
Repairing pages on the user's side predates modern AI. Richards and Hanson argued for transformation-based accessibility that adapts content in transit to individual abilities~\cite{richards2004broader}; Accessmonkey let users and developers share scripts that fix pages in the browser~\cite{bigham2007accessmonkey}; and Social Accessibility crowdsourced external metadata that assistive technology could consume without touching the original site~\cite{takagi2008social}. These systems achieved reliability by being narrow: each script or annotation encoded a specific, human-authored, human-verified fix, at the cost of coverage, since every new page needed new human effort. Coverage is exactly what a generative model promises to buy back; our work tests that trade and quantifies what it currently costs in reliability. A parallel line of work builds general-purpose web agents; benchmarks such as Mind2Web and WebArena show rapid progress but a persistent gap between open-weight and frontier models on realistic websites~\cite{deng2023mind2web,zhou2024webarena}, and none measures whether an agent's actions preserve or degrade accessibility. Recent systems explore this with modern models for specific tasks, for example restructuring cluttered e-commerce pages for screen reader users~\cite{yu2025cluttered} or generating alternative text for STEM images from a browser extension~\cite{pedemonte2025improving}. Our agent targets the complementary, presentation-level slice of the problem: type, contrast, spacing, focus visibility, and related style properties.

\subsection{LLMs for Web Accessibility}
A fast-growing literature applies LLMs to accessibility; a recent review finds most target text-centric issues, rely on large commercial models, and rarely evaluate with disabled users~\cite{aljedaani2026slr}. For remediation, Othman et~al.\ had ChatGPT repair 37 of 39 WAVE-detected violations on two sites~\cite{othman2023fostering}; at 88-site scale only about 70\% proved fixable, with systematic failures on contrast and perceivability~\cite{aljedaani2024chatgpt}. Detection has advanced faster than repair: generative pipelines outperform classical checkers~\cite{he2025enhancing}, LLMs can automate previously manual success criteria~\cite{lopezgil2025turning}, and yet model suggestions sometimes alter page semantics while claiming to fix them~\cite{delnevo2024interaction}. AccessGuru combines taxonomy-guided prompting with a multi-agent design and multiplies correction accuracy over naive prompting for HTML violations~\cite{fathallah2025accessguru}, evidence that architecture, not just model choice, drives repair quality. Generation is more sobering still: LLM-written interface code perpetuates barriers even with accessibility-oriented prompting~\cite{gurita2025barriers}, coding assistants need explicit scaffolding to help~\cite{mowar2025codea11y}, and the 2026 WebAIM Million names AI-assisted coding a likely contributor to the year's regression~\cite{webaim2026million}.

Our study differs on four axes at once: the intervention is user-side, running in the live browser rather than an authoring workflow; the models are small open-weight models running locally, a deliberate privacy constraint, rather than the frontier commercial models that dominate the literature~\cite{aljedaani2026slr}; the repair surface is CSS presentation rather than HTML structure or text; and, most distinctively, the protocol includes an accessible control condition designed to measure harm, not only benefit. To our knowledge no prior LLM remediation study reports a do-no-harm condition, though harm is precisely the risk that matters for a deployed agent.

\section{The Accessibility Agent}
\label{sec:system}
\begin{figure}
    \centering
    \includegraphics[width=\linewidth]{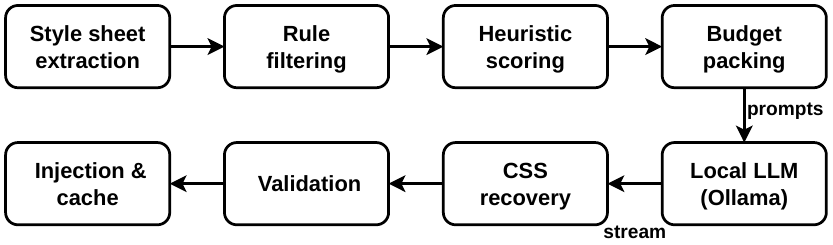}
    \caption{The agent pipeline. Page styling is extracted in the active tab, condensed to fit the model's context budget, sent with the accessibility prompts to a locally hosted model, and the recovered CSS is validated and injected as an additional adopted style sheet.}
    \label{fig:pipeline}
\end{figure}

The agent is a Manifest~V3 Chrome extension paired with an Ollama server hosting the model on the same machine~\cite{ollama2026}. Four goals shaped the design: no cooperation from site authors; local inference, since the agent sees every page the user visits; additive, reversible modifications; and consumer hardware, which in practice meant capping context at 10{,}000 tokens on a 16~GB GPU. Figure~\ref{fig:pipeline} shows the pipeline; each stage below corresponds to a component in the released source.

\subsection{Extraction}
On activation, the extension executes a script in the focused tab via the \texttt{chrome.scripting} API and walks \texttt{document.styleSheets}, collecting the text of every CSS rule. Cross-origin style sheets raise a security error on direct enumeration; for those the extension falls back to fetching the sheet by its URL and parsing the response. Production sites routinely ship megabytes of CSS, tens of thousands of tokens, so raw extraction alone far exceeds any local model's practical context.

\subsection{Filtering and Ranking}
\label{sec:ranking}
Context condensation proceeds in two passes. A filtering pass strips what costs tokens without carrying visual meaning: font import and animation rules; declarations whose value is \texttt{unset}, \texttt{initial}, or \texttt{revert} (utility frameworks use these to reset styling, not to style); and a curated list of properties set to \texttt{none}. Rules left empty are dropped.

A ranking pass then scores each surviving rule by its likely visual reach. Broad selectors (\texttt{*}, \texttt{html}, \texttt{body}) receive a large bonus and bare element selectors a smaller one, on the reasoning that they shape the most content; each occurrence of a high-impact property family (color, background, font, border, spacing, layout) adds to the score, as does \texttt{!important}, which usually marks a fix the site itself considered critical. Pseudo-element rules, hover and focus variants, and rules with very long selector lists (typically resets) are penalized. As a worked example, \texttt{body \{ font-family: ...; color: ...; \}} earns the broad-selector bonus plus one increment per impactful property and lands near the top of the ranking, while \texttt{.card:hover::after} with a decorative transition is stripped by the filter or ranked near the bottom. Rules are sorted by score and greedily packed until the character budget is exhausted. Because none of the tested models ships a tokenizer usable from a content script, token cost is approximated from character count as
\begin{equation}
\widehat{\text{tokens}} = \frac{\text{characters}}{2},
\end{equation}
a deliberately conservative ratio for token-dense CSS text, applied over a window of $\min(\text{model context}, 10{,}000)$ tokens with 40\% of the window reserved for the model's response. At the 10{,}000-token cap this admits roughly 20{,}000 characters of ranked CSS, which on the sites studied preserved the styling that dominates the visible page while discarding the long tail.

\subsection{Prompting}
Two prompts steer generation. The system prompt attempts to force raw CSS output and reads, verbatim:
\begin{quote}\footnotesize\ttfamily\sloppy
RESPOND ONLY WITH RAW CSS. NO WORDS. NO MARKDOWN. NO EXPLANATION. NO ANALYSIS. ONLY RESPOND WITH MODIFICATIONS. RESPONSE WILL BE APPLIED ADDITIVELY ON TOP OF EXISTING CSS. Output must start with a CSS selector. Output must end with a closing brace. Any response that contains English words outside of CSS values will be considered a failure.
\end{quote}
The user message contains the packed style sheet followed by an instruction to make additions that improve, where possible, the 18 metrics of Table~\ref{tab:metrics}, each stated with its quantitative threshold where one exists (for example, the WCAG~AA contrast ratios and the WCAG~2.2 minimum target size). The full user prompt is included in the replication package.

\subsection{Output Recovery, Validation, and Injection}
No model we tried, with any prompt phrasing, complied consistently with the raw-CSS instruction; responses arrived as bare CSS, fenced \texttt{css} blocks, fenced HTML with \texttt{style} elements, or unlabeled fences, often wrapped in prose. The agent therefore recovers CSS through a tolerant cascade: dedicated \texttt{css} fences anywhere in the response; then \texttt{style} elements harvested from fenced HTML via a DOM parser; then unlabeled fences; and finally the raw response itself, on the assumption the model obeyed. Recovered text is validated by parsing it into a \texttt{CSSStyleSheet}; responses that fail to parse or lack basic structure are rejected. Valid sheets are appended to the document's adopted style sheets, which layers them on top of the site's own CSS without modifying a byte of it: removing the adopted sheet restores the original page exactly, giving every intervention a one-step undo. Generated sheets are cached per URL in extension storage so revisits apply instantly without regeneration, with a user-facing toggle to disable or clear the cache, and privileged URL schemes (browser settings pages and the like) are excluded from injection. Chrome extensions can run this entire flow from a background service worker with no user action; given the output quality reported below, we kept a manual trigger instead, a decision we return to in Section~\ref{sec:discussion}.

\subsection{Implementation}
The agent comprises roughly 1{,}900 lines of TypeScript under Manifest~V3; the CSS filtering, ranking, budgeting, and recovery utilities are unit-tested pure functions, since they encode the study's context policy and must behave identically across researchers' machines. The full source and exact prompts ship in the replication package.

\section{Study Design}
\label{sec:study}

\subsection{Conditions and Websites}
We selected twenty live websites, listed in Table~\ref{tab:sites}. Ten form the \emph{sample} condition: high-traffic commercial sites (marketplaces, travel, media) chosen because inspection showed clear violations of the study metrics, so improvement is possible in principle. Ten form the \emph{control} condition: sites with strong accessibility reputations, including government digitalization portals and accessibility-focused organizations, judged to have negligible violations. The control condition operationalizes the do-no-harm hypothesis: on these sites an ideal agent should change little and break nothing, so any introduced violation is directly attributable to the intervention. The sample deliberately mixes English, Danish, and German sites and both international and regional operators, since a user-side agent must work wherever its user browses; CSS itself is language agnostic, but typographic and layout conventions vary across these markets, and an agent tuned on English-centric assumptions could fail differently elsewhere.

\begin{table}[t]
\caption{Study websites by condition.}
\label{tab:sites}
\begin{tabular}{@{}ll@{}}
\toprule
\textbf{Control (accessible)} & \textbf{Sample (violation-rich)} \\
\midrule
gov.uk & ticketmaster.dk \\
webaim.org & temu.com/dk \\
developer.mozilla.org & aliexpress.com \\
wikipedia.org & wish.com \\
bbc.com & amazon.de \\
dr.dk & ryanair.com/dk/da \\
borger.dk & forbes.com \\
digst.dk & nordic.ign.com \\
apple.com & fandom.com \\
microsoft.com & booking.com \\
\bottomrule
\end{tabular}
\end{table}

\subsection{Metrics}
\label{sec:metrics}
Sites were judged solely against an 18-item rubric (Table~\ref{tab:metrics}) assembled from WCAG~2.1 and 2.2 success criteria~\cite{wcag21,wcag22} and from the W3C COGA guidance on cognitive and learning disabilities~\cite{coga2021}, restricted to properties that CSS can in principle influence. The rubric spans perception (contrast, non-text contrast, color reliance), typography (font choice and size with respect to dyslexia and low vision, line height, wrapping), layout and reflow (320~px reflow, 200\% and 400\% zoom, content arrangement and hierarchy), and operation (touch target size, spacing of interactive elements, focus indicator visibility and contrast, cursor affordance). Structural concerns needing HTML or ARIA changes~\cite{aria12} were out of scope; Section~\ref{sec:future} returns to them.

\begin{table}[t]
\caption{Evaluation rubric: 18 CSS-addressable metrics.}
\label{tab:metrics}
\footnotesize
\begin{tabular}{@{}p{0.46\columnwidth}p{0.44\columnwidth}@{}}
\toprule
\textbf{WCAG-derived} & \textbf{COGA-derived} \\
\midrule
Text/background contrast (AA/AAA thresholds) & Font choice suited to dyslexia \\
Non-text contrast (3:1 minimum) & Font size for dyslexia and low vision \\
Meaning not conveyed by color alone & Content arrangement for orientation \\
Line height near 1.5$\times$ font size & Content hierarchy \\
Text wrapping behavior & Deliberate use of color symbolism \\
Reflow at 320~px without horizontal scroll & Spacing between interactive elements \\
200\% zoom without loss of function & Cursor styling signaling interactivity \\
400\% zoom (AAA) & \\
Touch target size (24$\times$24 px minimum) & \\
Spacing between interactive elements & \\
Visible focus indicators & \\
Focus indicator contrast (3:1) & \\
\bottomrule
\end{tabular}
\end{table}

\subsection{Models}
Six open-weight instruction-following models were drawn from the Ollama library, chosen to span parameter scales runnable on consumer GPUs and to vary vendor and architecture family: three smaller models, Qwen2.5-Coder~7B~\cite{hui2024qwen}, DeepSeek-R1~8B~\cite{deepseek2025r1}, and StarCoder2~7B~\cite{lozhkov2024starcoder2}, and three medium models, Gemma~3~12B~\cite{gemma2025}, Code~Llama~13B~\cite{roziere2023codellama}, and Phi-4~14B~\cite{abdin2024phi4}. All ran with Ollama's default quantized weights; selection favored availability and speed on the study hardware over expected capability (Section~\ref{sec:limits}). Varying vendor and family within each tier was deliberate: a shared failure across four independent lineages, which is what we observed, is harder to dismiss as one vendor's artifact. All models received identical prompts and the identical 10{,}000-token budget, so model identity is the only manipulated variable; per-model prompt tuning would confound exactly the comparison this design targets.

\subsection{Procedure and Scoring}
For each model and website, the evaluator loaded the site, captured a before screenshot, triggered the agent, and captured an after screenshot once the generated CSS was injected. Responses from which no valid CSS could be recovered were discarded and the agent rerun; after ten consecutive invalid responses the trial was recorded as producing no change and the protocol moved on. For every trial we logged the prompts, the qualified model name, the site, both screenshots, and the outcome. The outcome is the \emph{violation difference}: the count of rubric violations after minus before, so negative values indicate improvement. From the violation difference and the screenshots each trial was also categorized as an \emph{improvement} (at least one violation removed, none added on balance), a \emph{regression} (net violations added), a \emph{neutral edit} (visible change, no net effect on the rubric), or \emph{no change}. No change covers two protocol-identical cases, both scored zero: injected CSS with no observable effect, and trials abandoned after ten invalid responses; the log separates them cleanly only for StarCoder2, whose no-change column is entirely retry exhaustion, so elsewhere the category is an upper bound on genuine no-ops. Trials ran on the researchers' own consumer machines, GPUs in the 16~GB VRAM class, the deployment scenario the agent targets; the token budget was fixed at 10{,}000 for all models so every machine could run every condition identically, admitting roughly 20{,}000 characters of CSS, enough to include the large majority of ranked styling for every site. A single evaluator scored all trials; the absence of a second rater is a validity threat discussed in Section~\ref{sec:limits}.

\subsection{Analysis}
The original analysis plan tested each model's mean violation difference against zero with a two-tailed one-sample $t$ test per condition at $\alpha = .05$. We report those $p$ values as computed in the study and complement them with analyses matched to the data's granularity: the violation differences are small integers with many zeros, far from the $t$ test's assumptions at $n = 10$, so we add exact two-sided binomial sign tests of improvements against regressions (per model, per condition, and pooled) and Fisher exact tests comparing rates between conditions; sign tests discard magnitude but need no distributional assumptions. We also report the design's sensitivity: with $n = 10$ per cell, a two-tailed one-sample $t$ test at $\alpha = .05$ reaches 80\% power only for effects of about $|d| \geq 1.0$, so the study can rule out only very large effects and is best read as a structured pilot.

\section{Results}
\label{sec:results}

\subsection{Output Validity}
StarCoder2~7B failed completely: across all twenty websites it never produced a response from which the recovery cascade could extract valid CSS, so all twenty of its planned trials in the two conditions yielded no data beyond the failure itself. Every other model produced injectable CSS across the study, though frequently only via the fenced-block fallbacks rather than the demanded raw format; the trial log does not separately record isolated retry exhaustion for these models, which is why Section~\ref{sec:study} reads their no-change counts as upper bounds on genuine no-ops. Instruction compliance was thus a real cost even before considering repair quality: the strict system prompt of Section~\ref{sec:system} was violated routinely by every model.

\subsection{What the Models Changed}
\begin{figure*}[t]
\centering
\includegraphics[width=0.96\textwidth]{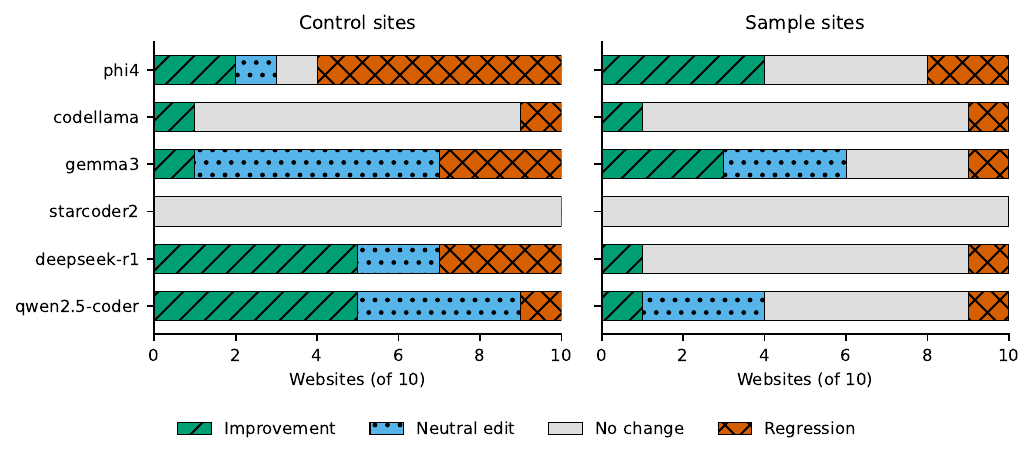}
\caption{Outcome categories per model and condition (10 websites each). StarCoder2 produced no valid CSS in any trial, so its bars record the resulting absence of modification. Improvements and regressions are roughly balanced on control sites; most sample-site trials changed nothing measurable.}
\Description{Two horizontal stacked bar charts, one for control sites and one for sample sites, with one bar per model showing counts out of ten websites split into improvement, neutral edit, no change, and regression. On control sites, qwen2.5-coder and deepseek-r1 each improved five sites, phi4 regressed six, codellama left eight unchanged, and starcoder2 shows ten no-change trials. On sample sites, phi4 improved four and regressed two, gemma3 improved three, and the other models mostly produced no change.}
\label{fig:changes}
\end{figure*}

Figure~\ref{fig:changes} and Table~\ref{tab:counts} summarize the categorized outcomes for the remaining five models, 50 trials per condition. Pooled over models, the control condition saw 14 improvements against 14 regressions (28\% each), and the sample condition 10 improvements against 6 regressions (20\% versus 12\%), with the remainder split between neutral edits and no visible change. Per-site violation differences were small in magnitude throughout, ranging from $-2$ to $+4$ on control sites and $-2$ to $+2$ on sample sites, and heavily zero-inflated: in 56 of these 100 trials the net violation count did not move at all. The single worst outcome in the study, four violations added to one page, occurred on a control site, and no trial removed more than two.

\begin{table}[t]
\caption{Categorized outcomes per model (I = improvement, N = neutral edit, U = unchanged, R = regression; 10 sites per cell).}
\label{tab:counts}
\begin{tabular}{@{}lcccc@{\hspace{9pt}}cccc@{}}
\toprule
 & \multicolumn{4}{c}{\textbf{Control}} & \multicolumn{4}{c}{\textbf{Sample}} \\
\cmidrule(r{7pt}){2-5}\cmidrule(l){6-9}
\textbf{Model} & I & N & U & R & I & N & U & R \\
\midrule
Qwen2.5-Coder 7B & 5 & 4 & 0 & 1 & 1 & 3 & 5 & 1 \\
DeepSeek-R1 8B   & 5 & 2 & 0 & 3 & 1 & 0 & 8 & 1 \\
StarCoder2 7B    & 0 & 0 & 10 & 0 & 0 & 0 & 10 & 0 \\
Gemma 3 12B      & 1 & 6 & 0 & 3 & 3 & 3 & 3 & 1 \\
Code Llama 13B   & 1 & 0 & 8 & 1 & 1 & 0 & 8 & 1 \\
Phi-4 14B        & 2 & 1 & 1 & 6 & 4 & 0 & 4 & 2 \\
\bottomrule
\end{tabular}
\end{table}

Model behavior diverged in style more than in efficacy. Qwen2.5-Coder and DeepSeek-R1 were the most active on control sites, each improving five of ten, but Qwen was far more conservative (one regression against DeepSeek's three) and both went largely inert on the sample sites, leaving most unchanged. Gemma~3 inverted the pattern: on control sites it produced mostly neutral edits plus three regressions and only one improvement, yet on sample sites it achieved three improvements against a single regression, the best sample-side balance. Code~Llama barely intervened anywhere, changing two sites per condition. Phi-4 had the highest variance: six control-site regressions, more than any other model, yet also the most sample-site improvements, four, and the most sample-site regressions. Across models, the improvements recorded in the trial log clustered heavily on the typographic metrics, font size and line height above all, which are exactly the metrics a single broad declaration can satisfy; metrics that demand element-specific judgment, such as user orientation and content arrangement, were both harder to score and essentially never improved. The regressions were collateral rather than gratuitous: in the trial records they arise from edits aimed at one metric that degraded conformance or usability elsewhere in the interface. Neutral edits, visible restyling with no rubric effect, occurred in 13 of 50 control trials but only 6 of 50 sample trials, so on the accessible sites the models were more active overall yet no more accurate.

\subsection{Hypothesis Tests}
Table~\ref{tab:stats} reports the original per-model $t$ tests together with our exact sign tests. Every $p$ value is far from significance: the $t$ tests range from $.62$ to $1.0$, and no per-model sign test falls below $.219$. Pooled across the five producing models, improvements and regressions are exactly balanced on control sites (14 versus 14, $p = 1.0$) and statistically indistinguishable on sample sites (10 versus 6, $p = .454$). Comparing conditions, improvement rates do not differ (Fisher exact $p = .483$); regressions were nominally 2.85 times more frequent on control than on sample sites, but this too does not reach significance ($p = .078$) and we flag it only as an exploratory trend. Each model's $p$ values are also similar across conditions, suggesting the intervention's footprint did not depend on the site being accessible or violation-rich. Neither directional hypothesis is supported: the study cannot conclude that the intervention improves violation-rich sites, and it cannot conclude that harm to accessible sites is rare. Given the $|d| \geq 1.0$ sensitivity bound, the honest summary is that any true effect of this intervention, positive or negative, is not large.

\begin{table}[t]
\caption{Per-model tests. $t$ tests are the study's two-tailed one-sample tests on violation differences; sign tests are exact two-sided binomial tests of improvements vs.\ regressions. StarCoder2 is omitted (no data).}
\label{tab:stats}
\begin{tabular}{@{}lcccc@{}}
\toprule
 & \multicolumn{2}{c}{\textbf{Control}} & \multicolumn{2}{c}{\textbf{Sample}} \\
\cmidrule(r{6pt}){2-3}\cmidrule(l){4-5}
\textbf{Model} & $p_t$ & $p_{\mathrm{sign}}$ & $p_t$ & $p_{\mathrm{sign}}$ \\
\midrule
Qwen2.5-Coder 7B & .95 & .219 & 1.0 & 1.0 \\
DeepSeek-R1 8B   & .83 & .727 & .90 & 1.0 \\
Gemma 3 12B      & .80 & .625 & .76 & .625 \\
Code Llama 13B   & 1.0 & 1.0  & .90 & 1.0 \\
Phi-4 14B        & .62 & .289 & .76 & .688 \\
\midrule
Pooled (sign)    & \multicolumn{2}{c}{$p = 1.0$} & \multicolumn{2}{c}{$p = .454$} \\
\bottomrule
\end{tabular}
\end{table}

\section{Discussion}
\label{sec:discussion}

\subsection{Answering the Research Questions}
\textbf{RQ1.} A zero-cooperation, privacy-preserving intervention is feasible: extraction, condensation, local generation, and additive injection all work, and could run fully in the background from a service worker. We kept the manual trigger, because autonomy is only a virtue when the actions are trustworthy, and Section~\ref{sec:results} shows they are not yet. \textbf{RQ2.} The rubric's metrics separate into three bands. Globally settable typography (font family, size, line height, wrapping) yields to one broad declaration, and these are precisely the metrics the models moved. Perception-dependent metrics (contrast, focus indicator contrast, color reliance) are CSS-settable in principle but require knowing which computed colors meet which elements, information absent from rule text, and were as often broken as fixed. Layout-emergent metrics (reflow, zoom robustness, target size, arrangement) depend on the whole cascade and were essentially never improved. The first band is a CSS-only agent's realistic near-term scope; the second needs perceptual grounding; the third likely needs structural change. \textbf{RQ3.} The intervention produced improvements and regressions at similar rates and no significant net effect in either condition. \textbf{RQ4.} Models differed in temperament, conservative Qwen and Code~Llama against volatile Phi-4, more than in achieved accessibility, and no between-model difference approaches significance at this sample size.

\subsection{Why Blind CSS Repair Fails}
The failure pattern is informative beyond the null result.
\paragraph{The model cannot see.} Contrast, reflow, target size, and focus visibility are properties of the \emph{rendered} page. A model shown only rule text must infer which foreground meets which background across cascade and inheritance. Small models guess, and a guessed color override is as likely to collide with a conformant element as to fix a failing one, matching prior findings that contrast is among the criteria LLMs handle worst~\cite{aljedaani2024chatgpt}.
\paragraph{Truncated context invites collateral damage.} Budget packing necessarily discards rules. Additive CSS injected on top of a partially observed cascade interacts with unseen selectors and specificity, so an edit that is locally sensible can be globally harmful. The exploratory asymmetry we observed, regressions concentrating on the accessible control sites, is consistent with the ceiling effect this mechanism predicts: tuned sites have the most conformant elements to break, while on violation-rich sites a blunt override can only tie or improve already failing metrics.
\paragraph{Instruction non-compliance is the norm.} Even the narrow contract of raw CSS output was breached by every model and met with total failure by one. Any deployed agent must treat model output as untrusted input to be parsed, validated, and sandboxed, never as a command to be executed verbatim.
\paragraph{Conformance is not experience.} Our rubric, like all guideline-derived instruments, bounds rather than measures lived accessibility~\cite{power2012guidelines,petrie2007relationship,vigo2013benchmarking}. Several neutral edits changed pages visibly without touching the rubric; whether such changes help or annoy real users is unknowable without user testing, which this study lacks. Finally, manual scoring of 18 metrics per capture is what limited this study to 120 trials; the missing component, an automated checker, is needed both as run-time guardrail and as experimental throughput, and Section~\ref{sec:instrument} delivers it.

\subsection{Implications for Assistive Agents}
For the community building assistive agents for real-world interfaces, we read these results as design requirements rather than as a verdict on the vision. First, \emph{perceptual grounding}: an agent judging contrast or reflow needs access to the rendered outcome, via screenshots, computed styles, or both, which argues for multimodal models or for pairing the LLM with deterministic computation of the very quantities WCAG defines numerically. Second, \emph{verification loops}: generation must be followed by automated checking before anything reaches the user's eyes. The browser makes this cheap: apply the candidate sheet, audit with an engine such as axe-core~\cite{axecore} plus computed-style probes of the numeric criteria, keep the sheet only if violations strictly decrease, otherwise roll back and reprompt with the findings. AccessGuru's gains from exactly this kind of structured, feedback-carrying pipeline on HTML repair point the same way~\cite{fathallah2025accessguru}. Third, \emph{minimal targeted diffs}: prompting for site-wide improvement over a truncated style dump invites exactly the global overrides we observed; an agent should instead localize one concrete violation, hand the model only the rules and computed styles relevant to the offending elements, and request the smallest edit that repairs it.

Beyond safety, targeting collapses the context problem, since a single violation's relevant styling fits comfortably in even a small model's window, making the token budgeting of Section~\ref{sec:ranking} a fallback rather than the default path. Fourth, \emph{reversibility and consent}: adopted style sheets, per-site caching, and a toggle made every change instantly revocable; given the observed error rates, silent autonomous modification would be irresponsible for now, and the human trigger is a feature, not a compromise. Fifth, \emph{context and locale awareness}: our trials spanned English, Danish, and German sites, and a CSS-level pipeline is anchored to none of them, a genuine advantage of operating below the language layer; but once an agent moves up to HTML, ARIA, and content, locale and cultural convention become first-class inputs, and evaluation samples must reflect that from the start. Finally, \emph{honesty about model scale}: privacy motivates local models, but at 7B to 14B parameters the models tested cannot yet carry this task alone; hybrid designs that reserve the LLM for what rules cannot do, and rules for what they can, are the plausible near-term path.

\section{From Diagnosis to Design: A Verified Repair Instrument}
\label{sec:instrument}
Acting on these requirements, we built and validated the verification infrastructure Study 1 lacked: a seeded-violation benchmark with exact ground truth, an automated auditor mirroring the CSS-decidable slice of the study rubric, and a repair loop whose accept rule makes regression on those checks impossible by construction. Everything here ran in a real Chromium instance; the numbers characterize the instrument itself, deliberately exercised with ground-truth and adversarial stub generators rather than any model, so its guarantees hold independently of model behavior.

\subsection{Benchmark and Auditor}
The benchmark comprises 42 locally served pages: per locale (English, Danish, German), ten pages carrying one to four seeded violations drawn from eight CSS-level seed types, plus four clean controls. Seeds cover low-contrast text and controls, color-only links, sub-12px body text, cramped line height, removed focus outlines, horizontal overflow at 320px, and sub-24px touch targets; ground truth records exactly which checks each page must fail. The auditor combines two axe-core rules (color contrast, link-in-text-block)~\cite{axecore} with five custom probes computed from real rendered styles and layout, including a cascade-aware scan for outline removal and a target-size check honoring the WCAG~2.2 inline-link exception; reflow is tested by physically resizing the viewport to 320px. Validation is exact: the auditor detects all 57 seeded violation instances, fires on zero of the twelve clean pages, and its fired rule set equals ground truth on every one of the 42 pages.

\subsection{The Verified Loop and Its Guarantee}
Per page the loop audits, requests CSS with the study's verbatim prompts (retries carry the concrete failing checks), validates that the response parses, injects, re-audits, and accepts only if total violations strictly decrease with no new violation type; otherwise it rolls back and retries, up to three attempts. Acceptance therefore cannot regress the automated checks. Table~\ref{tab:instrument} summarizes the validation. 
Against a generator that always emits harmful CSS, all 126 candidates were rejected and every page ended byte-identical to its start. Against a ground-truth fix generator, all 30 seeded pages were fully repaired, removing all 123 automated violations detected by the auditor. These correspond to the benchmark's 57 seeded ground-truth violations, since a single seeded violation may trigger multiple automated checks. Every edit to a clean page was correctly rejected for lacking measurable benefit; a mixed generator cycling fixes, harm, and invalid prose reached the same end state through retries. The loop also carries the multimodal path the discussion calls for, capturing a page screenshot and attaching it to requests for vision-capable models such as Gemma~3, exercised end to end apart from inference itself.

\begin{table}[t]
\caption{Instrument validation in real Chromium (stub generators; characterizes the harness, not any model).}
\label{tab:instrument}
\begin{tabular}{@{}lr@{}}
\toprule
Seeded violations detected & 57/57 \\
Clean pages with any false positive & 0/12 \\
Pages with fired rules $=$ ground truth & 42/42 \\
Harmful candidates rejected (adversarial stub) & 126/126 \\
Regressions after adversarial run & 0 \\
Seeded pages fully repaired (ground-truth stub) & 30/30 \\
\bottomrule
\end{tabular}
\end{table}

\subsection{What the Instrument Enables}
The instrument converts both halves of the problem. At run time it is the guardrail Section~\ref{sec:discussion} argues any deployed agent needs, with residual risk confined to regressions its checks cannot see, which is why human spot-checks of accepted repairs remain in the protocol. As evaluation infrastructure it replaces weeks of manual scoring with unattended runs: a model's full-benchmark evaluation is one command and a few inference-bound hours, ground truth removes live-site drift, and the locale structure supports the analysis of linguistic variation that real-world assistive agents require. Running the six study models, in text and vision conditions, is the immediate next step; those model results are not reported here. The benchmark, auditor, loop, and validation logs ship in the replication package as reusable evaluation and safety infrastructure for community efforts deploying assistive agents.

\section{Limitations and Threats to Validity}
\label{sec:limits}
The trial count is small: ten sites per condition per model, 120 planned trials including the 100 for the five models that produced injectable CSS during the study. Sensitivity is bounded at roughly $|d| = 1.0$, and a single scoring error moves observed rates by several points. A single evaluator scored everything, with no second rater or reliability statistic, and several rubric items (content arrangement, hierarchy, color symbolism) require judgment. No automated checker cross-validated the manual scores, and no users from the target populations evaluated the modified pages, so nothing here speaks to experienced accessibility. Live commercial sites are moving targets; content variation between the before and after captures cannot be fully excluded. Only home pages were tested, and each model and site pair rests on a single recorded generation, so sampling variance is unmeasured; a different draw could flip individual trials. Finally, results are specific to the quantized builds, the 10{,}000-token cap, and the single prompt pair; larger models or different prompting could change the picture, though no prompt variant we tried during development eliminated the problems reported.

\section{Future Directions}
\label{sec:future}
Three extensions follow directly. Extending the repair surface from CSS to HTML and ARIA attributes would address the structural barriers, such as accessible names and roles, that matter most to screen reader users~\cite{aria12}; the architecture requires no change beyond the extraction and injection targets, but the token economics worsen, sharpening the need for targeted diffs. The immediate step is running the study models, text and vision alike, through the instrument of Section~\ref{sec:instrument}, whose throughput makes powered comparisons and locale-stratified analysis routine; extending its checks toward fuller rubric coverage, with human sampling against automated tools' blind spots~\cite{vigo2013benchmarking,ara2025inclusive}, strengthens both guardrail and science. And user studies with the intended populations, which this project attempted but could not recruit for in time, remain the only way to connect check movement to lived benefit. Personalization is a further avenue: the same pipeline could load per-user metric profiles, for example prioritizing dyslexia-friendly typography for one user and enlarged touch targets for another, which is precisely where a user-side agent holds a structural advantage over author-side remediation.

\section{Conclusion}
We set out to test whether small local LLMs can repair web accessibility from the user's side, and ended up delivering the three pieces the question requires. The agent works; that part is solved engineering, released in full. The measurement is honest: under a protocol that weighs harm as carefully as benefit, unverified generation improved and regressed live pages at similar rates, fixing what a global declaration can fix and breaking what it cannot see. And the missing component is now built: a validated benchmark and verified repair loop that catches every seeded violation, provably rejects harmful edits, and turns a week of manual evaluation into an unattended run. The lesson is plain. An accessibility agent earns trust not by guessing better but by verifying every edit against the rendered page before a user sees it, and that infrastructure now exists for anyone pursuing a web that adapts to its users when its authors do not.

\section*{Data Availability}
The replication package contains the extension source (the exact study snapshot preserved alongside a tested release build), the verbatim prompts, the rubric and procedure, the summary trial data, the seeded benchmark with ground truth, the verified-loop harness with its validation logs and a one-command reproduction of Table~\ref{tab:instrument}, and the analysis scripts. The live-site trial-level records and screenshots are retained by the authors and available on request; the per-model $t$ tests are reported as computed in the study, and the package recomputes them automatically once the trial-level export is supplied.

\bibliographystyle{ACM-Reference-Format}
\bibliography{references}

@misc{webaim2026million,
  author       = {{WebAIM}},
  title        = {The {WebAIM} Million: The 2026 Report on the Accessibility of the Top 1,000,000 Home Pages},
  year         = {2026},
  howpublished = {\url{https://webaim.org/projects/million/}},
  note         = {Accessed July 2026}
}

@misc{wcag21,
  author       = {Andrew Kirkpatrick and Joshue O'Connor and Alastair Campbell and Michael Cooper},
  title        = {Web Content Accessibility Guidelines ({WCAG}) 2.1},
  year         = {2018},
  howpublished = {W3C Recommendation},
  note         = {\url{https://www.w3.org/TR/WCAG21/}}
}

@misc{wcag22,
  author       = {{W3C}},
  title        = {Web Content Accessibility Guidelines ({WCAG}) 2.2},
  year         = {2023},
  howpublished = {W3C Recommendation},
  note         = {\url{https://www.w3.org/TR/WCAG22/}}
}

@misc{coga2021,
  author       = {{W3C Web Accessibility Initiative}},
  title        = {Making Content Usable for People with Cognitive and Learning Disabilities},
  year         = {2021},
  howpublished = {W3C Working Group Note},
  note         = {\url{https://www.w3.org/TR/coga-usable/}}
}

@misc{aria12,
  author       = {{W3C}},
  title        = {Accessible Rich Internet Applications ({WAI-ARIA}) 1.2},
  year         = {2023},
  howpublished = {W3C Recommendation},
  note         = {\url{https://www.w3.org/TR/wai-aria-1.2/}}
}

@inproceedings{vigo2013benchmarking,
  author    = {Markel Vigo and Justin Brown and Vivienne Conway},
  title     = {Benchmarking Web Accessibility Evaluation Tools: Measuring the Harm of Sole Reliance on Automated Tests},
  booktitle = {Proceedings of the 10th International Cross-Disciplinary Conference on Web Accessibility (W4A '13)},
  year      = {2013},
  publisher = {ACM}
}

@inproceedings{power2012guidelines,
  author    = {Christopher Power and Andr{\'e} Freire and Helen Petrie and David Swallow},
  title     = {Guidelines Are Only Half of the Story: Accessibility Problems Encountered by Blind Users on the Web},
  booktitle = {Proceedings of the SIGCHI Conference on Human Factors in Computing Systems (CHI '12)},
  year      = {2012},
  pages     = {433--442},
  publisher = {ACM}
}

@inproceedings{petrie2007relationship,
  author    = {Helen Petrie and Omar Kheir},
  title     = {The Relationship between Accessibility and Usability of Websites},
  booktitle = {Proceedings of the SIGCHI Conference on Human Factors in Computing Systems (CHI '07)},
  year      = {2007},
  pages     = {397--406},
  publisher = {ACM}
}

@article{ara2025inclusive,
  author  = {Jinat Ara and Cecilia Sik-Lanyi and Arpad Kelemen and Tibor Guzsvinecz},
  title   = {An Inclusive Framework for Automated Web Content Accessibility Evaluation},
  journal = {Universal Access in the Information Society},
  volume  = {24},
  number  = {2},
  pages   = {1581--1607},
  year    = {2025}
}

@article{fakrudeen2025evaluation,
  author  = {Mohammed Fakrudeen},
  title   = {Evaluation of the Accessibility and Usability of University Websites: A Comparative Study of the Gulf Region},
  journal = {Universal Access in the Information Society},
  volume  = {24},
  number  = {2},
  pages   = {1883--1898},
  year    = {2025}
}

@inproceedings{richards2004broader,
  author    = {John T. Richards and Vicki L. Hanson},
  title     = {Web Accessibility: A Broader View},
  booktitle = {Proceedings of the 13th International Conference on World Wide Web (WWW '04)},
  year      = {2004},
  pages     = {72--79},
  publisher = {ACM}
}

@inproceedings{bigham2007accessmonkey,
  author    = {Jeffrey P. Bigham and Richard E. Ladner},
  title     = {Accessmonkey: A Collaborative Scripting Framework for Web Users and Developers},
  booktitle = {Proceedings of the 2007 International Cross-Disciplinary Conference on Web Accessibility (W4A '07)},
  year      = {2007},
  publisher = {ACM}
}

@inproceedings{takagi2008social,
  author    = {Hironobu Takagi and Shinya Kawanaka and Masatomo Kobayashi and Takashi Itoh and Chieko Asakawa},
  title     = {Social Accessibility: Achieving Accessibility through Collaborative Metadata Authoring},
  booktitle = {Proceedings of the 10th International ACM SIGACCESS Conference on Computers and Accessibility (ASSETS '08)},
  year      = {2008},
  publisher = {ACM}
}

@inproceedings{othman2023fostering,
  author    = {Achraf Othman and Amira Dhouib and Aljazi {Nasser Al Jabor}},
  title     = {Fostering Websites Accessibility: A Case Study on the Use of the Large Language Models {ChatGPT} for Automatic Remediation},
  booktitle = {Proceedings of the 16th International Conference on PErvasive Technologies Related to Assistive Environments (PETRA '23)},
  year      = {2023},
  pages     = {707--713},
  publisher = {ACM},
  doi       = {10.1145/3594806.3596542}
}

@inproceedings{aljedaani2024chatgpt,
  author    = {Wajdi Aljedaani and Abdulrahman Habib and Ahmed Aljohani and Marcelo Eler and Yunhe Feng},
  title     = {Does {ChatGPT} Generate Accessible Code? Investigating Accessibility Challenges in {LLM}-Generated Source Code},
  booktitle = {Proceedings of the 21st International Web for All Conference (W4A '24)},
  year      = {2024},
  pages     = {165--176},
  publisher = {ACM},
  doi       = {10.1145/3677846.3677854}
}

@inproceedings{fathallah2025accessguru,
  author    = {Nadeen Fathallah and Daniel Hern{\'a}ndez and Steffen Staab},
  title     = {{AccessGuru}: Leveraging {LLMs} to Detect and Correct Web Accessibility Violations in {HTML} Code},
  booktitle = {Proceedings of the 27th International ACM SIGACCESS Conference on Computers and Accessibility (ASSETS '25)},
  year      = {2025},
  publisher = {ACM}
}

@inproceedings{gurita2025barriers,
  author    = {Alexandra-Elena Gurita and Radu-Daniel Vatavu},
  title     = {When {LLM}-Generated Code Perpetuates User Interface Accessibility Barriers, How Can We Break the Cycle?},
  booktitle = {Proceedings of the 22nd International Web for All Conference (W4A '25)},
  year      = {2025},
  pages     = {124--134},
  publisher = {ACM}
}

@article{lopezgil2025turning,
  author  = {Juan-Miguel L{\'o}pez-Gil and Juanan Pereira},
  title   = {Turning Manual Web Accessibility Success Criteria into Automatic: An {LLM}-Based Approach},
  journal = {Universal Access in the Information Society},
  volume  = {24},
  number  = {1},
  pages   = {837--852},
  year    = {2025},
  doi     = {10.1007/s10209-024-01108-z}
}

@article{pedemonte2025improving,
  author  = {Giacomo Pedemonte and Maurizio Leotta and Marina Ribaudo},
  title   = {Improving Web Accessibility With an {LLM}-Based Tool: A Preliminary Evaluation for {STEM} Images},
  journal = {IEEE Access},
  volume  = {13},
  pages   = {107566--107582},
  year    = {2025}
}

@inproceedings{delnevo2024interaction,
  author    = {Giovanni Delnevo and Manuel Andruccioli and Silvia Mirri},
  title     = {On the Interaction with Large Language Models for Web Accessibility: Implications and Challenges},
  booktitle = {Proceedings of the 2024 IEEE 21st Consumer Communications \& Networking Conference (CCNC)},
  year      = {2024},
  pages     = {1--6},
  publisher = {IEEE}
}

@article{he2025enhancing,
  author  = {Ziyao He and Syed Fatiul Huq and Sam Malek},
  title   = {Enhancing Web Accessibility: Automated Detection of Issues with Generative {AI}},
  journal = {Proceedings of the ACM on Software Engineering},
  volume  = {2},
  number  = {FSE},
  pages   = {2264--2287},
  year    = {2025}
}

@inproceedings{yu2025cluttered,
  author    = {Yaman Yu and Bektur Ryskeldiev and Ayaka Tsutsui and Matthew Gillingham and Yang Wang},
  title     = {From Cluttered to Clear: Improving the Web Accessibility Design for Screen Reader Users in E-commerce with Generative {AI}},
  booktitle = {Proceedings of the 27th International ACM SIGACCESS Conference on Computers and Accessibility (ASSETS '25)},
  year      = {2025},
  publisher = {ACM}
}

@inproceedings{mowar2025codea11y,
  author    = {Peya Mowar and Yi-Hao Peng and Jason Wu and Aaron Steinfeld and Jeffrey P. Bigham},
  title     = {{CodeA11y}: Making {AI} Coding Assistants Useful for Accessible Web Development},
  booktitle = {Proceedings of the 2025 CHI Conference on Human Factors in Computing Systems (CHI '25)},
  year      = {2025},
  publisher = {ACM}
}

@inproceedings{aljedaani2026slr,
  author    = {Wajdi Aljedaani and Rubel Hassan Mollik},
  title     = {Large Language Models for Web Accessibility: A Systematic Literature Review},
  booktitle = {Proceedings of the 23rd International Web for All Conference (W4A '26)},
  year      = {2026},
  publisher = {ACM},
  doi       = {10.1145/3800424.3800452}
}

@inproceedings{deng2023mind2web,
  author    = {Xiang Deng and Yu Gu and Boyuan Zheng and Shijie Chen and Samuel Stevens and Boshi Wang and Huan Sun and Yu Su},
  title     = {Mind2Web: Towards a Generalist Agent for the Web},
  booktitle = {Advances in Neural Information Processing Systems 36 (NeurIPS 2023), Datasets and Benchmarks Track},
  year      = {2023}
}

@inproceedings{zhou2024webarena,
  author    = {Shuyan Zhou and Frank F. Xu and Hao Zhu and Xuhui Zhou and Robert Lo and Abishek Sridhar and Xianyi Cheng and Tianyue Ou and Yonatan Bisk and Daniel Fried and Uri Alon and Graham Neubig},
  title     = {WebArena: A Realistic Web Environment for Building Autonomous Agents},
  booktitle = {Proceedings of the Twelfth International Conference on Learning Representations (ICLR 2024)},
  year      = {2024}
}

@misc{hui2024qwen,
  author = {Hui, Binyuan and Yang, Jian and Cui, Zeyu and Yang, Jiaxi and Liu, Dayiheng and Zhang, Lei and Liu, Tianyu and Zhang, Jiajun and Yu, Bowen and Lu, Keming and others},
  title  = {Qwen2.5-Coder Technical Report},
  year   = {2024},
  note   = {arXiv:2409.12186}
}

@misc{deepseek2025r1,
  author = {Guo, Daya and Yang, Dejian and Zhang, Haowei and Song, Junxiao and Wang, Peiyi and Zhu, Qihao and Xu, Runxin and Zhang, Ruoyu and Ma, Shirong and Bi, Xiao and others},
  title  = {DeepSeek-R1: Incentivizing Reasoning Capability in {LLMs} via Reinforcement Learning},
  year   = {2025},
  note   = {arXiv:2501.12948}
}

@misc{gemma2025,
  author = {{Gemma Team}},
  title  = {Gemma 3 Technical Report},
  year   = {2025},
  note   = {arXiv:2503.19786}
}

@misc{roziere2023codellama,
  author = {Baptiste Rozi{\`e}re and others},
  title  = {Code Llama: Open Foundation Models for Code},
  year   = {2023},
  note   = {arXiv:2308.12950}
}

@misc{abdin2024phi4,
  author = {Abdin, Marah and Aneja, Jyoti and Behl, Harkirat and Bubeck, S{\'e}bastien and Eldan, Ronen and Gunasekar, Suriya and Harrison, Michael and Hewett, Russell J and Javaheripi, Mojan and Kauffmann, Piero and others},
  title  = {Phi-4 Technical Report},
  year   = {2024},
  note   = {arXiv:2412.08905}
}

@misc{lozhkov2024starcoder2,
  author = {Lozhkov, Anton and Li, Raymond and Allal, Loubna Ben and Cassano, Federico and Lamy-Poirier, Joel and Tazi, Nouamane and Tang, Ao and Pykhtar, Dmytro and Liu, Jiawei and Wei, Yuxiang and others},
  title  = {StarCoder 2 and The Stack v2: The Next Generation},
  year   = {2024},
  note   = {arXiv:2402.19173}
}

@misc{ollama2026,
  author       = {{Ollama}},
  title        = {Ollama: Run Large Language Models Locally},
  year         = {2026},
  howpublished = {\url{https://ollama.com}},
  note         = {Accessed July 2026}
}

@misc{axecore,
  author       = {{Deque Systems}},
  title        = {axe-core: Accessibility Engine for Automated Web {UI} Testing},
  year         = {2026},
  howpublished = {\url{https://github.com/dequelabs/axe-core}},
  note         = {Accessed July 2026}
}

\end{document}